\documentclass{article}

\usepackage{arxiv}

\usepackage[utf8]{inputenc} 
\usepackage[T1]{fontenc}    
\usepackage{hyperref}       
\usepackage{url}            
\usepackage{booktabs}       
\usepackage{amsfonts}       
\usepackage{nicefrac}       
\usepackage{microtype}      
\usepackage{lipsum}
\usepackage{graphicx}
\graphicspath{{fig/}}
\usepackage{amssymb,bm,cite,empheq,here,mathtools,times}
\usepackage[dvipsnames]{xcolor}

\newtheorem{defn}{Definition}
\newtheorem{thm}{Theorem}

\usepackage{algorithmic}
\usepackage{algorithm}

\title{Steady State Covariance Steering via\\ Sparse Intervention}

\author{
 Yosuke Inoue and Masaki Inoue\\
  Department of Applied Physics and Physico-Informatics,\\
  Keio University, \\
  \texttt{yo\_hirofuku@keio.jp, minoue.z6@keio.jp} \\
}

\begin{document}
\maketitle
\begin{abstract}
This paper addresses the steady state covariance steering for linear dynamical systems via structural intervention on the system matrix. 
We formulate the covariance steering problem as the minimization of the Kullback-Leibler (KL) divergence between the steady state and target Gaussian distributions. 
To solve the problem, we develop a solution method, hereafter referred to as the proximal gradient-based algorithm, of promoting sparsity in the structural intervention by integrating the objective into a proximal gradient framework with $L_1$ regularization. 
The main contribution of this paper lies in the analytical expression of the KL divergence gradient with respect to the intervention matrix: the gradient is characterized by the solutions to two Lyapunov equations related to the state covariance equation and its adjoint.
Numerical simulations demonstrate that the proximal gradient-based algorithm effectively identifies sparse, structurally-constrained interventions to achieve precise covariance steering.
\end{abstract}


\section{Introduction}
\label{sec:intro}

Various real-world challenges of ranging from medical treatment to traffic and pedestrian flow stabilization, can be viewed as the control of state distributions. 
Motivated by such applications, 
covariance steering for dynamical systems, along with its variants such as covariance control, has been extensively studied in the literature \cite{hotz1985,collins1987,hsieh1990,fujioka1994,chen2016,bakolas2018,balci2022,kashima2024,liu2025}. 
Early works include state feedback covariance steering \cite{hotz1985,collins1987} and output feedback approaches with controller parameterization \cite{hsieh1990,fujioka1994}. 
In addition, covariance steering over a finite horizon has remained an active area within the control community in recent years \cite{chen2016,bakolas2018,balci2022,kashima2024,liu2025}.

This paper addresses covariance steering via direct intervention to the system matrix under specified structural constraints such as sparsity.  To this end, we formulate the covariance steering problem as the minimization of the Kullback-Leibler (KL) divergence between the steady state and reference Gaussian distributions.  Then, we propose a solution algorithm that integrates gradient descent with $L_1$ regularization.  The algorithm ensures the sparsity of the intervention matrix, which is essential for addressing practical limitations in large-scale complex systems.
The main theoretical contribution is the analytical expression of the KL divergence gradient with respect to the intervention matrix: the gradient is characterized by the solutions to two Lyapunov equations, the state covariance equation and its adjoint, providing a scalable solution for high-dimensional systems.
\smallskip

\noindent
\textbf{Notation:}\
Symbol $\mathbb{R}$ denotes the set of real numbers.
For a random variable $x$, $x \sim P$ means it follows the probability distribution $P$.
$I_n$ denotes the $n \times n$ identity matrix.
$0_n$ denotes the $n$-dimensional zero vector.
For a matrix $M$, $\operatorname{tr}(M)$ denotes its trace.
$\|M\|_{\mathrm{fro}}$ denotes the Frobenius norm of $M$. 
$\|M\|_1 = \sum_{i,j} |M_{ij}|$ represents the entry-wise $L_1$ norm.
For matrices $M_1$ and $M_2$, $\langle M_1, M_2 \rangle = \operatorname{tr}(M_1^\top M_2)$ denotes the Frobenius inner product.



\section{Problem Formulation}
\label{sec:formulation}

We consider the linear dynamical system under intervention, described by
\begin{align}
    \label{eq:lssU}
    x(k+1) = (A + U)\,x(k) + B\,w(k),
\end{align}
where $x\in \mathbb{R}^{n}$ denote the state, $w\in \mathbb{R}^{m}$ denote the disturbance satisfying $w(k) \sim \Ncal(0_m, I_m)$, $A\in \mathbb{R}^{n \times n}$ and $B\in \mathbb{R}^{n \times m}$ are constant system matrices, and $U \in \Ucal \subseteq \mathbb{R}^{n \times n}$ is the \textit{intervention matrix}, representing the parameters that are manipulable by the system designer.
Supposing that $A+U$ is Schur stable, the state variable $x(k)$ asymptotically follows a zero-mean Gaussian distribution as well.
In other words, letting $\Sigma_U$ denote the state covariance matrix that depends on $U$, we have
$x(k) \sim \Ncal(0_n, \Sigma_U)$, $k\to \infty$.

Let $\Vcal$ denote the set of the intervention candidates. 
Then, the set of admissible intervention matrices is defined as
\begin{align}
    \label{eq:U_subject}
    \Ucal = 
    \left\{ 
    U \in \mathbb{R}^{n \times n} \;\middle|\; U_{pq}= 0, (p,q) \notin \Vcal, \|U\|_1 \leq m
    \right\}.
\end{align}
The constraint, $U_{pq} = 0,(p,q) \notin \Vcal$, represents a sparsity requirement, indicating that the $(p,q)$ element of $U$ is inaccessible. 
This corresponds to selectively intervening in a predetermined subset indexed by $\mathcal{V}$.
The constraint, $\|U\|_1 \leq m$, promotes the sparsity in $U$ by limiting its cardinality.

We then formulate the problem of steady state covariance steering.  
To measure the difference between the state covariance matrix $\Sigma_U$ and and the reference one $\Sigma_{\mathrm{ref}}$, we employ the KL divergence.
\begin{defn} 
    \label{deff:KL} (KL divergence)
    For a random variable $x$, the KL divergence between the two probability distributions $P$ and $Q$ is defined as 
    \begin{align}
        \label{eq:deff_KL}
        \text{KL}(P||Q) := \int p(x) \log \frac{p(x)}{q(x)} dx,
    \end{align}
    where $p$ and $q$ denote the probability density functions of $P$ and $Q$, respectively.
\end{defn}

To simplify the notation, we further let $\operatorname{KL}(\Sigma_1 \| \Sigma_2)$ denote the KL divergence between two different zero-mean Gaussian distributions $\Ncal(0_n,\Sigma_1)$ and $\Ncal(0_n,\Sigma_2)$.  Then, the following proposition on the KL divergence holds.
\begin{prop} 
    \label{prop:KL}
    For an $n$-dimensional random variable, 
    the KL divergence between the two zero-mean Gaussian distributions $\mathcal{N}(0_n, \Sigma_1)$ and $\mathcal{N}(0_n, \Sigma_2)$ is expressed as 
    \begin{align}
        \label{eq:KL_gaus}
        \text{KL}(\Sigma_1 || \Sigma_2) = \frac{1}{2} \left( \operatorname{tr}(\Sigma_2^{-1} \Sigma_1) - n + \ln\frac{|\Sigma_2|}{|\Sigma_1|} \right).
    \end{align}
\end{prop}
Noting that the steady state covariance is characterized by the solution to the Lyapunov equation (see e.g. \cite{collins1987}), we formulate the steady state covariance steering as follows.

\begin{prob}
    \label{prob:covariance}
    Given $\Sigma_{\mathrm{ref}}$, solve the following optimization problem to find $U$:
    \begin{subequations}
        \label{eq:prob_st}
        \begin{align}
            \label{subeq:prob_st}
            \underset{U \in \mathcal{U},\Sigma \geq 0}{\mathrm{min}} \quad & J(\Sigma) = \operatorname{KL}(\Sigma \| \Sigma_\mathrm{ref}) \\
            \label{subeq:prob_st_rst1}
            \mathrm{s.t.} \quad & (A+U) \Sigma (A+U)^\top - \Sigma + B B^\top = 0.
        \end{align}
    \end{subequations}
\end{prob} 

Typical applications of Problem \ref{prob:covariance} include medical interventions at the genetic level \cite{inoue2026,shen2023} and advanced traffic management strategies, such as signal control and congestion tolling\cite{kamal2015,siri2021freeway}.
In such control problems, the control action is modeled as a direct intervention in the system matrix $A$, representing structural changes in the underlying dynamics.

Covariance steering for linear dynamical systems has been extensively studied in the literature \cite{hotz1985,collins1987,hsieh1990,fujioka1994,chen2016,bakolas2018,balci2022,kashima2024,liu2025}. 
In contrast to prior works that focus on steering using \textit{dynamic} control, this paper adopts a \textit{static control} and aims at the steady-state distribution.
As stated in Problem \ref{prob:covariance}, this paper addresses an optimization problem subject to the Lyapunov equation \eqref{subeq:prob_st_rst1}, which serves as a nonlinear constraint with respect to the intervention matrix $U$. 
Solving such a constrained optimization problem, often interpreted as optimization over a manifold, is non-trivial and presents a significant computational challenge, as the solution is not analytically tractable.

\section{Proximal Gradient-based  Algorithm}
\label{sec:algo}

\subsection{Gradient Characterization}
\label{subsec:gradient}

We provide a characterization of the gradient $d J/d U$ on the Lyapunov equation constraint.
We employ the adjoint method\cite{yan2016} for efficient gradient evaluation. 
We have the following theorem.

\begin{thm}
    \label{theo:grad_MICS}
    The gradient of $J$ with respect to $U$, subject to the Lyapunov constraint defined in (\ref{subeq:prob_st_rst1}), is given by
    \begin{align}
        \label{eq:grad_MICS}
        \left.\frac{d J}{d U}\right|_{U = U^\dagger} = 2 \Lambda^\dagger (A+U^\dagger) \Sigma^\dagger,
    \end{align}
    where $\Sigma^\dagger$ is the solution $\Sigma$ to the primal Lyapunov equation
    \begin{align}
        \label{eq:grad_MICS_sigma}
        \left(A+U^\dagger\right) \Sigma \left(A+U^\dagger\right)^\top - \Sigma + B B^\top = 0,
    \end{align}
    and $\Lambda^\dagger$ is the solution $\Lambda$ to the adjoint Lyapunov equation
    \begin{align}
        \label{eq:grad_MICS_lambda}
        \left(A+U^\dagger\right)^\top \Lambda \left(A+U^\dagger\right) - \Lambda + \left.\frac{\partial J}{\partial \Sigma}\right|_{\Sigma = \Sigma^\dagger} = 0.
    \end{align}
\end{thm}

\begin{proo}  
To derive the gradient ${d J}/{d U}$, we employ the adjoint method.  
To this end, we define the Lagrangian $\mathcal{L}$ by incorporating the constraint using the matrix adjoint variable $\Lambda$:
\begin{align}
    \label{eq:lagrangian}
    &\Lcal(U, \Sigma, \Lambda) \notag \\
    &= J(\Sigma) + \langle \Lambda, (A+U) \Sigma (A+U)^\top - \Sigma + B B^\top \rangle.
\end{align}

The derivation of the gradient is characterized by three stationarity conditions: (i) the primal Lyapunov equation is satisfied by the stationarity condition $\partial \mathcal{L} / \partial \Lambda = 0$. (i\hspace{-0.1em}i) the adjoint variable $\Lambda$ is uniquely determined by the stationarity condition $\partial \mathcal{L} / \partial \Sigma = 0$. (i\hspace{-0.1em}i\hspace{-0.1em}i) the total derivative ${d J}/{d U}$ is reduced to the partial derivative $\partial \mathcal{L} / \partial U$, presenting the explicit gradient expression.
The details are given as follows.

(i) We first determine $\Sigma$ such that the stationarity condition $\partial \Lcal / \partial \Lambda = 0$ is satisfied, which is equivalent to 
(\ref{eq:grad_MICS_sigma}). 
Under the condition, 
we see that $\Lcal(U, \Sigma, \Lambda) = J(\Sigma)$ holds, which implies the equality
\begin{align} 
    \label{eq:equiv_total_deriv} 
    \frac{d \Lcal}{d U} = \frac{dJ}{d U}. 
\end{align}

(i\hspace{-0.1em}i)
We next determine $\Lambda$ such that the stationarity condition $\partial \Lcal / \partial \Sigma = 0$ is satisfied, which corresponds to (\ref{eq:grad_MICS_lambda}).
According to the chain rule, 
the total derivative of $\Lcal$ is expressed as
\begin{align}
    \label{eq:grad_reduction}
    \frac{d \Lcal}{dU} 
    = \frac{\partial \Lcal}{\partial U} + \frac{\partial \Lcal}{\partial \Sigma} \frac{\partial \Sigma}{\partial U}
    = \frac{\partial \Lcal}{\partial U}.
\end{align}

(i\hspace{-0.1em}i\hspace{-0.1em}i)
Finally, from (\ref{eq:equiv_total_deriv}) and (\ref{eq:grad_reduction}), we show that the gradient is expressed by (\ref{eq:grad_MICS}).  
The step-by-step derivation is as follows.
For notational simplicity, we let $A_U := A+U$. 
Then, the identity ${\partial A_U}/{\partial U} = I_n$ holds.
Consequently, it holds that 
\begin{align}
    \label{eq:dLdU_intro}
    \frac{dJ}{d U}=
    \frac{\partial \Lcal}{\partial U} = \frac{\partial \Lcal}{\partial A_U} \frac{\partial A_U}{\partial U} = \frac{\partial \Lcal}{\partial A_U}.
\end{align}
By focusing on the term containing $A_U$ and using the cyclic property of the trace, we have:
\begin{align}
    \label{eq:dldau}
    \frac{\partial \Lcal}{\partial A_U} &= \frac{\partial}{\partial A_U} \operatorname{tr} \left( A_U \Sigma A_U^\top \Lambda^\top \right)\notag\\
    & = \Lambda^\top A_U \Sigma + \Lambda A_U \Sigma^\top.
\end{align}
Recall that both $\Sigma$ and $\Lambda$ are the solutions to Lyapunov equations, which are inherently symmetric. 
Then, we have the expression
\begin{align}
    \label{eq:dldau_simplified}
    \frac{\partial \Lcal}{\partial A_U} = 2 \Lambda A_U \Sigma.
\end{align}

By evaluating the primal stationarity condition (i) at $U=U^\dagger$ to find $\Sigma = \Sigma^\dagger$, and subsequently solving the adjoint stationarity condition (i\hspace{-0.1em}i) at $U=U^\dagger$ and 
$\Sigma=\Sigma^\dagger$ to find $\Lambda = \Lambda^\dagger$,
we have the gradient expression in Theorem \ref{theo:grad_MICS}.
\hfill \qed
\end{proo}

\subsection{Optimization Algorithm}
\label{subsec:update}

Having derived the analytical gradient, we now provide an overview of the solution algorithm for Problem \ref{prob:covariance} based on the proximal gradient method.
Letting $l$ denote the iteration index, each iteration consists of the following two steps.
\begin{itemize}
    \item[1.] Gradient Step \\
    The gradient $dJ/dU$ is calculated at current $U^{(l)}$ according to \rtheo{theo:grad_MICS}. 
    Using this gradient, an intermediate value $V^{(l+1)}$ is obtained.
    \item[2.] Proximal Step \\
    Applying the soft-thresholding function to $V^{(l+1)}$ yields the updated control input $U^{(l+1)}$.
    This proximal step accounts for the $L_1$ regularization term within the optimization process.
\end{itemize}

Specifically, the optimization algorithm is described by 
\begin{subequations}
    \label{eq:update_MICS}
    \begin{align}
        \label{eq:update_MICS_V}
        V^{(l+1)} &= U^{(l)} - \eta \left.\frac{d J}{d U} \right|_{U = U^{(l)}}, \\
        \label{eq:update_MICS_U}
        U^{(l+1)} &= \text{prox}_{\eta \lambda} \left(V^{(l+1)}\right),
    \end{align}
\end{subequations}
where $\eta$ denotes the learning rate and $\lambda$ is the regularization parameter. 
In addition, $\operatorname{prox}_{\eta \lambda}(\cdot)$ represents the proximal operator, which is defined as follows (see e.g. the book \cite{nagahara2020sparsity} for details).
\begin{defn}
\label{deff:prox_matrix}
    (Proximal operator)
    The proximal operator for a matrix $V^{(l)}$ associated with the entry-wise $L_1$ norm $\|\cdot\|_1$ is defined as
    \begin{align}
        \label{eq:prox_matrix_def}
        \operatorname{prox}_{\eta \lambda}\left(V^{(l)}\right) = \mathop{\argmin}_{U} \left( \frac{1}{2} \|V^{(l)} - U\|_\mathrm{fro}^2 + \eta \lambda \|U\|_1 \right).
    \end{align}
\end{defn}
In this definition, the threshold is scaled by the learning rate $\eta$ following the standard proximal gradient descent framework.
This operator corresponds to the soft-thresholding function for the $L_1$ regularization term.
In addition, it is known that the proximal operator can be computed entry-wise and admits the following analytical form as
\begin{align}
    \label{eq:prox_entry}
    \operatorname{prox}_{\eta \lambda} \left(V^{(l)}_{ij}\right) =
        \begin{cases}
          V^{(l)}_{ij} - \eta \lambda,  & \text{if } ~ V^{(l)}_{ij} > \eta \lambda, \\
          0, & \text{if } \left| V^{(l)}_{ij}\right| \leq \eta \lambda, \\
          V^{(l)}_{ij} + \eta \lambda,  & \text{if } ~ V^{(l)}_{ij} < - \eta\lambda.
        \end{cases}
\end{align}
Based on the formulation in \eqref{eq:prox_entry}, we discuss the relationship between the proximal operator and $L_1$ regularization. 
First, focusing on the middle case of the right-hand side in \eqref{eq:prox_entry}, it implies that if the absolute value of $V^{(l)}_{ij}$ is less than or equal to $\eta \lambda$ (i.e., a sufficiently small value), $V^{(l)}_{ij}$ is truncated to zero. 
Next, as seen in the upper and lower cases, when the absolute value of $V^{(l)}_{ij}$ exceeds $\eta \lambda$, the operator does not return the original value but instead shrinks it toward the origin by $\eta \lambda$.
This operation is widely known as the soft-thresholding operator, which is the proximal mapping of the $L_1$ norm.
By forcing values within the threshold $[-\eta\lambda, \eta\lambda]$ to be exactly zero, this operator directly promotes a sparse solution, which is the primary objective of $L_1$ regularization.

Finally, the the proximal gradient-based algorithm is summarized in \ralgo{algo:MICS}.

\begin{algorithm}[t]
	\caption{Proximal Gradient-based Algorithm}
	\label{algo:MICS}
	\begin{algorithmic}[1]    
	\REQUIRE Reference covariance matrix $\Sigma_{\mathrm{ref}}$, system matrices $A, B$, learning rate $\eta$, regularization parameter $\lambda$, threshold $\epsilon$, initial guess $U^{(0)}$
    \ENSURE Local optimizer $U$
	\STATE $l \gets 1$
    \WHILE{$\left|d J/d U\right|_{\mathrm{fro}} \geq \epsilon$}
	    \STATE Solve \req{eq:grad_MICS_sigma} to obtain $\Sigma^{(l)}$
        \STATE Solve \req{eq:grad_MICS_lambda} to obtain $\Lambda^{(l)}$
		\STATE Calculate $dJ/dU$ via \req{eq:grad_MICS} with $\Sigma^{(l)}$ and $\Lambda^{(l)}$
		\STATE Calculate $V^{(l)}$ via \req{eq:update_MICS_V}
        \STATE Calculate $U^{(l)}$ via \req{eq:update_MICS_U}
		\STATE $l \gets l + 1$
    \ENDWHILE
	\RETURN $U$
	\end{algorithmic}
\end{algorithm}


\section{Numerical Simulation}
\label{sec:simu}

In this section, we verify the effectiveness of the proximal gradient-based algorithm through a numerical simulation.

The system to be controlled is described by the state equation (\ref{eq:lssU}), 
where the system matrices are given by
\begin{align*}
    A &= \begin{bmatrix}
    0.386 & 0 & 0 & 0 & 0.161 \\
    0 & 0.461 & 0 & -0.047 & 0 \\
    -0.042& 0 & 0.317 & 0.134 & -0.117\\
    0 & 0 & 0.134 & 0.401 & -0.157\\
    0.161 & 0 & -0.117& -0.157 & 0.85
    \end{bmatrix}, \\
    B &= \sqrt{0.003} I_5.
\end{align*}
The reference of the state covariance matrix $\Sigma_{\mathrm{ref}}$ is given by
\begin{align*}
    \Sigma_{\mathrm{ref}} = 
    \begin{bmatrix}
    0.0025 & 0.001 & 0.0002 & -0.0014 & -0.0002 \\
    0.001 & 0.0027 & -0.0003 & -0.0014 & 0.0002 \\
    0.0002 & -0.0003 & 0.0013 & 0.0006 & -0.0003 \\
    -0.0014 & -0.0014 & 0.0006 & 0.0093 & -0.0004 \\
    -0.0002 & 0.0002 & -0.0003 & -0.0004 & 0.0012
    \end{bmatrix}.
\end{align*}
The intervention matrix $U$ must satisfy the constraints in \req{eq:U_subject} with $m = 4$.

Through the simulation, we examine whether the algorithm can steer the covariance matrix $\Sigma_U$ of the state $x$ toward the target covariance matrix $\Sigma_{\mathrm{ref}}$.
Furthermore, we evaluate the effectiveness of the $L_1$ regularization in promoting sparsity of the intervention matrix and investigate the relationship between the steering performance and the degree of sparsity by varying the regularization parameter $\lambda$.
Additional settings required to run the algorithm are as follows:
\begin{itemize}
\item The initial intervention matrix $U^{(0)}$ was set such that it contains a single, small, non-zero element at the (5,5) position.

\item The maximum number of iterations for the algorithm was set to $100$.

\item The learning rate and the regularization parameter in \ralgo{algo:MICS} were set to $\eta = 0.1$ and $\lambda = 0.5$, respectively.
\end{itemize}

\begin{figure}[t]
    \centering
    \includegraphics[width=0.8\linewidth]{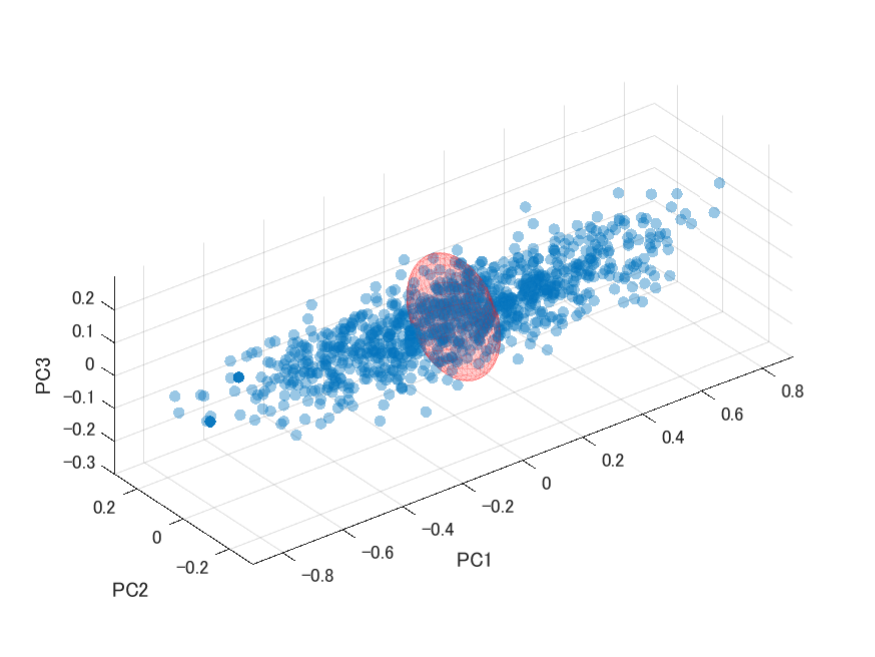}
    \caption{Covariance ellipsoid of the reference covariance matrix $\Sigma_{\mathrm{ref}}$ and scatter plot of $x$ projected onto the principal component space (Without control).}
    \label{fig:scatter_without_control}
\end{figure}

\begin{figure}[t]
    \centering
    \includegraphics[width=0.8\linewidth]{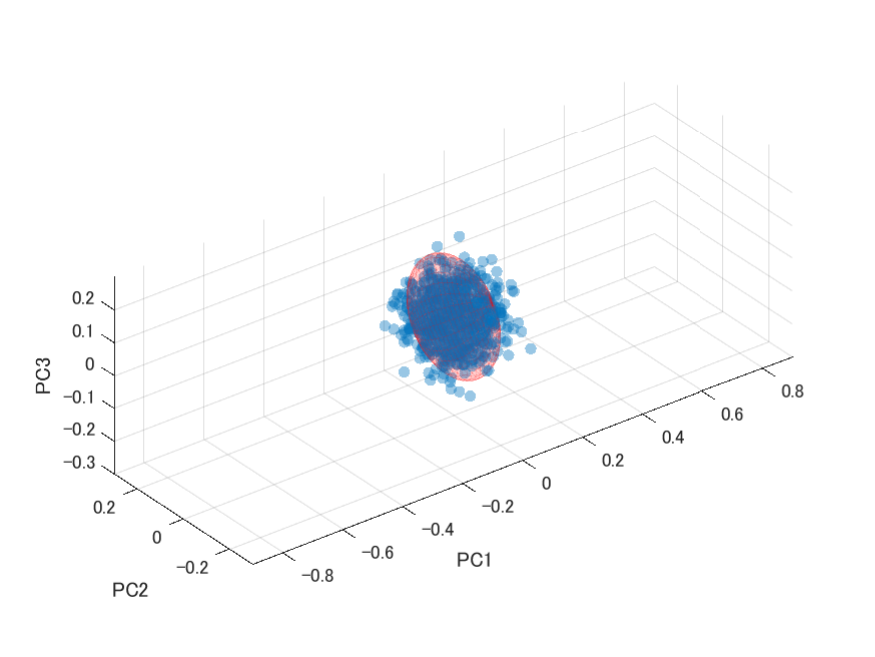}
    \caption{Covariance ellipsoid of the reference covariance matrix $\Sigma_{\mathrm{ref}}$ and scatter plot of $x$ projected onto the principal component space (With control).}
    \label{fig:scatter_with_control}
\end{figure}

We present the results of the simulation.
\rfig{fig:scatter_without_control} and \rfig{fig:scatter_with_control} show scatter plots of $x(k)$ at $k=50$ without and with control by the control, respectively.
In both \rfig{fig:scatter_without_control} and \rfig{fig:scatter_with_control}, The red ellipsoid represents the covariance ellipsoid corresponding to the reference covariance matrix $\Sigma_{\mathrm{ref}}$.
The ellipsoid is drawn to contain 99\% of the points sampled from the distribution $x \sim \Ncal(0_5,\Sigma_{\mathrm{ref}})$.
The ellipse is obtained by performing principal component analysis (PCA) on the dataset without control and visualizing it in a three-dimensional principal component space using the first, second and third principal components.
Each scatter plot in the figures shows the state $x(k) \in \mathbb{R}^5$ at time $k=50$ projected onto the same three-dimensional principal component space.
In \rfig{fig:scatter_without_control}, the points spread widely along the PC1 axis and deviate substantially from the reference covariance ellipsoid.
In contrast, in \rfig{fig:scatter_with_control}, almost all points are concentrated within the target covariance ellipsoid.

\begin{figure}[t]
    \centering
    \includegraphics[width=0.8\linewidth]{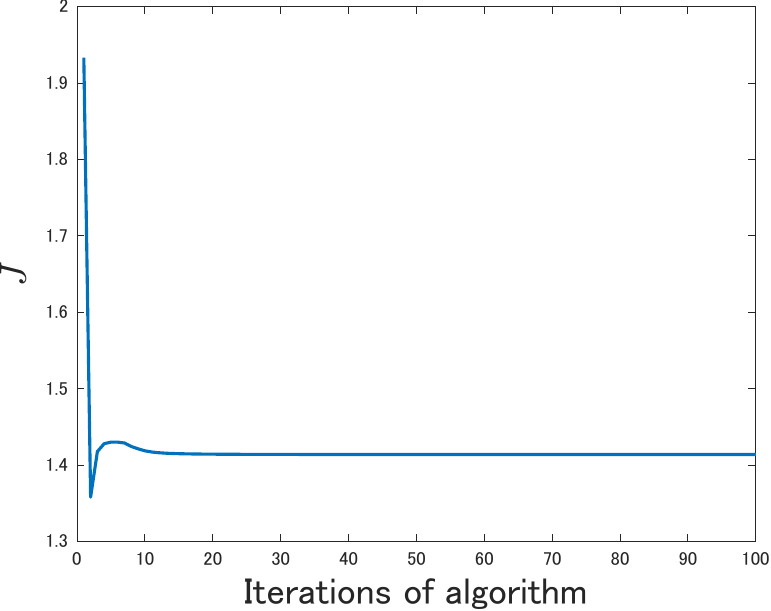}
    \caption{Evolution of $J$ against the number of iterations.}
    \label{fig:iteration_J}
\end{figure}

The control results are quantitatively evaluated based on the value of $J$. 
The evolution of $J$ is shown in \rfig{fig:iteration_J}.
As shown in \rfig{fig:iteration_J}, the objective $J$ converges to approximately $1.4$ as the iterations proceed.

At the termination of the algorithm, the obtained control input $U$ had non-zero values only at the following entries:$U_{11} = -0.0425$, $U_{33} = -0.0825$, $U_{51} = -0.0539$, $U_{55} = -0.6897$,
which indeed satisfies $U \in \mathcal{U}$.

\begin{figure}[t]
    \centering
    \includegraphics[width=0.8\linewidth]{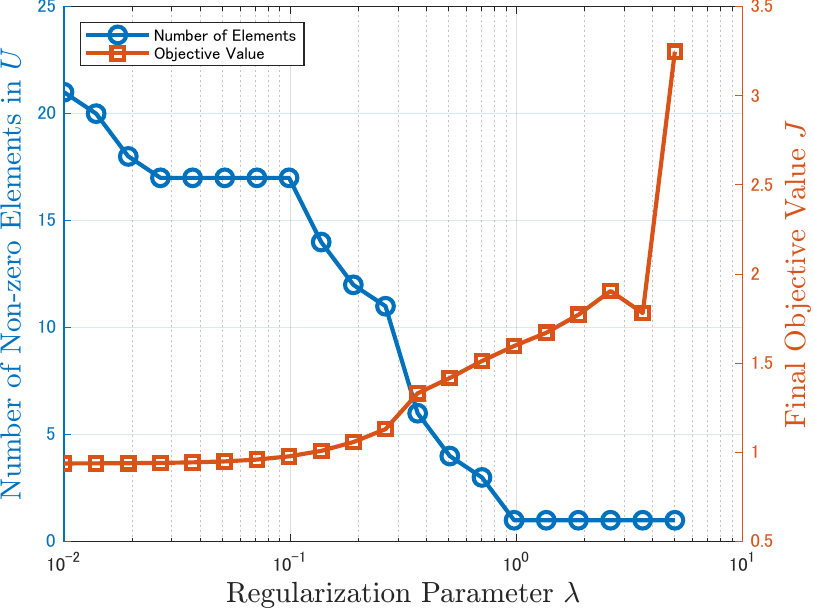}
    \caption{Relationship between the regularization parameter $\lambda$, the number of non-zero elements in the intervention matrix $U$ (blue plot), and the final objective value $J$ (orange plot).}
    \label{fig:lambda_num_J}
\end{figure}

\rfig{fig:lambda_num_J} shows the relationship between the regularization parameter $\lambda$, the number of non-zero elements in the intervention matrix $U$, and the final objective value $J$.
It can be observed that as $\lambda$ increases, the number of non-zero elements decreases at the cost of an increase in the objective value $J$, illustrating a clear trade-off between the sparsity of the intervention and the control performance.

These results demonstrate that the proximal gradient-based algorithm successfully achieves covariance steering while preserving the sparsity of the solution. 
Furthermore, it is observed that the degree of sparsity can be tuned by adjusting the regularization parameter, revealing a clear trade-off between the sparsity of the intervention and the control performance.

\section{Conclusion}
\label{sec:conclusion}

This paper presented a steady state covariance steering algorithm for linear dynamical systems through sparse structural intervention. 
The main contribution is the analytical characterization of the gradient using two Lyapunov equations, which enables an $L_1$-regularized proximal gradient algorithm to identify minimal structural interventions.
Future research will apply the proposed algorithm to healthcare and/or transportation systems where structural sparsity is critical. 
Additionally, we aim to extend the theory to induce instability in specific directions, investigating the conditions under which sparse interventions can effectively destabilize stable systems to trigger desired state transitions.


\bibliographystyle{plain}        
\bibliography{ref}    
\end{document}